# Multiloop Calculations in the String-Inspired Formalism: The Single Spinor-Loop in QED

Michael G. Schmidt [*]

*Institut für Theoretische Physik*
*Universität Heidelberg*
*Philosophenweg 16*
*D-69120 Heidelberg, Germany*

Christian Schubert [†]

*Institut für Hochenergiephysik Zeuthen*
*DESY Deutsches Elektronen-Synchrotron*
*Platanenallee 6*
*D-15738 Zeuthen, Germany*

## Abstract

We use the worldline path-integral approach to the Bern-Kosower formalism for developing a new algorithm for calculation of the sum of all diagrams with one spinor loop and fixed numbers of external and internal photons. The method is based on worldline supersymmetry, and on the construction of generalized worldline Green functions. The two-loop QED $\beta$ – function is calculated as an example.

---

[*] e-mail address M.G.Schmidt@ThPhys.Uni-Heidelberg.de
[†] e-mail address schubert@hades.ifh.de

**1. Introduction**

In 1992, Bern and Kosower [1] used string theory to derive a new formalism for the calculation of one-loop amplitudes in ordinary quantum field theory which is equivalent to Feynman diagrams [2], but leads to a significant reduction of the number of terms to be computed in gauge theory calculations. This property was then successfully exploited to obtain both five-point gluon [3] and four-point graviton amplitudes [4].

Following this, Strassler [5] showed that, at least for scalar and spinor loops, the resulting integral representations can be derived in a more elementary way. In this approach, one writes the one-loop effective action as a (super)particle path-integral, and evaluates this path-integral in analogy to the Polyakov path integral, i.e. using worldline Green functions appropriate to a one-dimensional field theory on the circle.

This reformulation turned out to be useful for various calculations of one-loop effective actions [6, 7, 8] and yielded, in particular, a new method for the calculation of the inverse mass expansion which is nonrecursive, manifestly gauge invariant, and suitable to computerization [7, 9].

Progress has been made along different lines to generalize the Bern-Kosower formalism beyond one loop, using methods either based on the calculation of higher genus string amplitudes [10], on the use of a separate worldline path integral for every internal propagator [11], or on a stringlike reorganization of standard Feynman parameter integrals [12]. A Hamiltonian approach has also been considered [13].

Recently, we have proposed [14] a multiloop generalization of Strassler's approach, based on the concept of worldline Green functions for multiloop diagrams. Those Green functions have been explicitly constructed for the general two-loop graph, and for a loop with an arbitrary number of propagator insertions. If used with global proper-time variables, this allows to derive integral representations combining whole classes of Feynman diagrams into compact expressions.

While knowledge of the worldline Green functions is, in principle, sufficient to treat arbitrary scalar diagrams, more work has to be done to obtain the final integral representations for multiloop amplitudes in general quantum field theories.

In the present paper, we take up the study of quantum electrodynamics, and consider a class of amplitudes which is the simplest one for our purpose, namely the N–photon amplitude with a single spinor loop.

## 2. One-Loop Amplitudes

First let us shortly review how one-loop calculations are done in this formalism [5, 7, 14]. For calculation of the one-loop effective action induced by a massive spinor-loop in a background gauge field, one would start with the following worldline path integral representation [15, 16, 17, 18, 19]:

$$\begin{aligned}\Gamma[A] &= -2\int_0^\infty \frac{dT}{T} e^{-m^2 T} \mathrm{tr} \int \mathcal{D}x \mathcal{D}\psi \\ &\quad \times \exp\left[-\int_0^T d\tau \left(\frac{1}{4}\dot{x}^2 + \frac{1}{2}\psi\dot{\psi} + ieA_\mu \dot{x}^\mu - ie\psi^\mu F_{\mu\nu}\psi^\nu\right)\right]\end{aligned} \tag{1}$$

Here the $x^\mu(\tau)$'s are the periodic functions from the circle with circumference $T$ into $D$ – dimensional spacetime, and the $\psi^\mu(\tau)$'s their antiperiodic Grassmannian supersymmetric partners. In the nonabelian case – which we will not consider in this paper – path ordering would be implied.

For calculation of the effective action, one first splits the coordinate path integral into center of mass and relative coordinates,

$$\begin{aligned}\int \mathcal{D}x &= \int dx_0 \int \mathcal{D}y \\ x^\mu(\tau) &= x_0^\mu + y^\mu(\tau) \\ \int_0^T d\tau\, y^\mu(\tau) &= 0 \ .\end{aligned} \tag{2}$$

One then Taylor-expands the external field at $x_0$, and evaluates the path integrals over $y$ and $\psi$ by Wick contractions, as in a one-dimensional field theory on the circle. The Green functions to be used are those adapted to the (anti-) periodicity conditions,

$$\begin{aligned}\langle y^\mu(\tau_1) y^\nu(\tau_2)\rangle &= -g^{\mu\nu} G_B(\tau_1, \tau_2) = -g^{\mu\nu}\left[|\tau_1 - \tau_2| - \frac{(\tau_1-\tau_2)^2}{T}\right], \\ \langle \psi^\mu(\tau_1)\psi^\nu(\tau_2)\rangle &= \frac{1}{2} g^{\mu\nu} G_F(\tau_1, \tau_2) = \frac{1}{2} g^{\mu\nu} \mathrm{sign}(\tau_1-\tau_2) \ .\end{aligned} \tag{3}$$

Our normalization is such that for the free path integrals

$$\int \mathcal{D}y \exp\left[-\int_0^T d\tau \frac{1}{4}\dot{y}^2\right] = [4\pi T]^{-\frac{D}{2}}$$
$$\int \mathcal{D}\psi \exp\left[-\int_0^T d\tau \frac{1}{2}\psi\dot{\psi}\right] = 1 \quad .$$
(4)

The result of this evaluation is the one-loop effective Lagrangian $\mathcal{L}(x_0)$.

One-loop scattering amplitudes are obtained by specializing to a background consisting of a finite number of plane waves. This amounts to the same thing as defining integrated vertex operators

$$\int_0^T d\tau \left[\dot{x}^\mu \varepsilon_\mu - 2i\psi^\mu \psi^\nu k_\mu \varepsilon_\nu\right]\exp[ikx(\tau)] \tag{5}$$

for external photons of definite momentum and polarization, and calculating multiple insertions of those vertex operators into the free path integral.

Using the worldline superfield formalism of [22, 18], this calculus may be cast into manifestly supersymmetric form. Eq.( 1) then becomes

$$\Gamma[A] = -2\int_0^\infty \frac{dT}{T} e^{-m^2 T} \int \mathcal{D}X \exp^{-\int_0^T d\tau \int d\theta \left[-\frac{1}{4}XD^3X - ieDX^\mu A_\mu(X)\right]}, \tag{6}$$

where

$$X^\mu = x^\mu + \sqrt{2}\,\theta\psi^\mu$$
$$D = \frac{\partial}{\partial\theta} - \theta\frac{\partial}{\partial\tau}$$
$$\int d\theta\,\theta = 1 \quad .$$
(7)

The photon vertex operator is rewritten as

$$-\int_0^T d\tau d\theta \varepsilon_\mu DX^\mu \exp[ikX], \tag{8}$$

and the worldline propagators may be combined into a superpropagator

$$\hat{G}(\tau_1,\theta_1;\tau_2,\theta_2) = G_B(\tau_1,\tau_2) + \theta_1\theta_2 G_F(\tau_1,\tau_2). \tag{9}$$

From this superfield formalism, it is not difficult to derive the following important "substitution rule" [1, 5, 6]:

Evaluation of the bosonic path integral in general leads to an expression consisting of an exponential factor $\exp\left[\sum_{i<j} G_B(\tau_i, \tau_j) p_i p_j\right]$, multiplied by a polynomial in the first and second derivatives of $G_B$,

$$\begin{aligned} \dot{G}_B(\tau_1, \tau_2) &= \text{sign}(\tau_1 - \tau_2) - 2\frac{(\tau_1 - \tau_2)}{T} \\ \ddot{G}_B(\tau_1, \tau_2) &= 2\delta(\tau_1 - \tau_2) - \frac{2}{T} \end{aligned} \tag{10}$$

(here and in the following, a "dot" denotes differentiation with respect to the first variable). The $\ddot{G}_B$'s can always be eliminated by partial integrations on the worldline, and once this has been done, all contributions from fermionic Wick contractions may be taken into account by replacing every closed cycle of $\dot{G}_B$'s appearing, say $\dot{G}_B(\tau_{i_1}, \tau_{i_2}) \dot{G}_B(\tau_{i_2}, \tau_{i_3}) \cdots \dot{G}_B(\tau_{i_n}, \tau_{i_1})$, by its supersymmetrization:

$$\begin{aligned} \dot{G}_B(\tau_{i_1}, \tau_{i_2}) \dot{G}_B(\tau_{i_2}, \tau_{i_3}) \cdots \dot{G}_B(\tau_{i_n}, \tau_{i_1}) &\to \dot{G}_B(\tau_{i_1}, \tau_{i_2}) \dot{G}_B(\tau_{i_2}, \tau_{i_3}) \cdots \dot{G}_B(\tau_{i_n}, \tau_{i_1}) \\ &- G_F(\tau_{i_1}, \tau_{i_2}) G_F(\tau_{i_2}, \tau_{i_3}) \cdots G_F(\tau_{i_n}, \tau_{i_1}). \end{aligned} \tag{11}$$

Note that the result would vanish if supersymmetry was not broken by the boundary conditions. This substitution rule effectively replaces the calculation of Dirac traces.

### 3. Scalar Multiloop Amplitudes

In our treatment of the two-loop scalar diagram [14], the starting point had been to insert into the free one-loop scalar path integral

$$\int_0^\infty \frac{dT}{T} e^{-m^2 T} \mathcal{D}x \exp\left[-\int_0^T d\tau \frac{\dot{x}^2}{4}\right], \tag{12}$$

a scalar propagator

$$\int_0^T d\tau_a \int_0^T d\tau_b \langle \phi(x(\tau_a)) \phi(x(\tau_b)) \rangle. \tag{13}$$

This propagator was then written in the Schwinger proper-time representation,

$$\langle \phi(x(\tau_a)) \phi(x(\tau_b)) \rangle = \int_0^\infty d\bar{T} e^{-m^2 \bar{T}} (4\pi\bar{T})^{-\frac{D}{2}} \exp\left[-\frac{(x(\tau_a) - x(\tau_b))^2}{4\bar{T}}\right]. \tag{14}$$

The exponent was considered part of the worldline lagrangian for the loop path integral, and absorbed into the bosonic worldline Green function. This results in a modified Green function

$$G_B^{(1)}(\tau_1,\tau_2) = G_B(\tau_1,\tau_2) + \frac{1}{2}\frac{[G_B(\tau_1,\tau_a) - G_B(\tau_1,\tau_b)][G_B(\tau_2,\tau_a) - G_B(\tau_2,\tau_b)]}{\bar{T} + G_B(\tau_a,\tau_b)} \quad (15)$$

valid for Wick contractions of operators inserted into the loop path integral.

The procedure generalizes to the case of $m$ propagator insertions, and leads to modified Green functions

$$\begin{aligned} G_B^{(m)}(\tau_1,\tau_2) &= G_B(\tau_1,\tau_2) \\ &+ \frac{1}{2}\sum_{k,l=1}^{m} [G_B(\tau_1,\tau_{a_k}) - G_B(\tau_1,\tau_{b_k})]A_{kl}^{(m)}[G_B(\tau_2,\tau_{a_l}) - G_B(\tau_2,\tau_{b_l})], \end{aligned} \quad (16)$$

with a symmetric $m \times m$ – matrix $A^{(m)}$ defined by

$$\begin{aligned} A^{(m)} &= \left[\bar{T} - \frac{B}{2}\right]^{-1} \\ \bar{T}_{kl} &= \bar{T}_k \delta_{kl} \\ B_{kl} &= G_B(\tau_{a_k},\tau_{a_l}) - G_B(\tau_{a_k},\tau_{b_l}) - G_B(\tau_{b_k},\tau_{a_l}) + G_B(\tau_{b_k},\tau_{b_l}) \end{aligned} \quad (17)$$

($\bar{T}_1,\ldots\bar{T}_m$ are the proper-time variables for the inserted propagators).

**4. QED Multiloop Amplitudes**

For scalar electrodynamics, it is obvious what should replace the propagator insertion eq.( 13). In this case, the one-loop path-integral eq.( 1) reduces to

$$\Gamma[A] = \int_0^\infty \frac{dT}{T} e^{-m^2 T} \int \mathcal{D}x \exp\left[-\int_0^T d\tau \left(\frac{1}{4}\dot{x}^2 + ieA_\mu \dot{x}^\mu\right)\right] \quad (18)$$

(note the deletion of the global factor of $-2$, which takes care of statistics and degrees of freedom). This expression may also be interpreted as a Wilson loop expectation value. It is well-known, however (see e.g. [18, 20, 21]), that the first-order correction to a scalar Wilson loop (due to exchange of one internal photon) may be written in terms of a worldline current-current interaction,

$$-\frac{e^2}{2}\frac{\Gamma(\lambda)}{4\pi^{\lambda+1}}\int_0^T d\tau_a \int_0^T d\tau_b \frac{\dot{x}^\mu(\tau_a)\dot{x}_\mu(\tau_b)}{((x(\tau_a)-x(\tau_b))^2)^\lambda} \quad , \tag{19}$$

with $\lambda = \frac{D}{2} - 1$.

For scalar electrodynamics, we will therefore insert one copy of this expression into the path integral eq.( 18) for every internal photon. The denominators will be written in the proper-time representation eq.( 14), which leads to the same generalized bosonic two-loop worldline Green function as in the scalar case; the numerators will remain, and participate in the Wick contractions.

As in the one-loop case, the transition to spinor electrodynamics may then be accomplished by supersymmetrization, which replaces eq.( 19) by

$$\frac{e^2}{2}\frac{\Gamma(\lambda)}{4\pi^{\lambda+1}}\int_0^T d\tau_a d\theta_a \int_0^T d\tau_b d\theta_b \frac{DX_a^\mu DX_{b\mu}}{((X_a-X_b)^2)^\lambda} \quad . \tag{20}$$

In components the double integral reads, after a bit of algebra,

$$\int_0^T d\tau_a \int_0^T d\tau_b \left\{ -\frac{\dot{x}_a^\mu \dot{x}_{b\mu}}{((x_a-x_b)^2)^\lambda} - 4\lambda \frac{(x_a^\mu - x_b^\mu)(\psi_b^\mu \psi_b^\nu \dot{x}_{a\nu} - \psi_a^\mu \psi_a^\nu \dot{x}_{b\nu})}{((x_a-x_b)^2)^{\lambda+1}} \right.$$
$$\left. +8\lambda \frac{(\psi_a^\mu \psi_{b\mu})^2}{((x_a-x_b)^2)^{\lambda+1}} - 16\lambda(\lambda+1)\frac{(x_a^\mu - x_b^\mu)(x_a^\nu - x_b^\nu)\psi_{a\mu}\psi_{b\nu}\psi_a^\kappa \psi_{b\kappa}}{((x_a-x_b)^2)^{\lambda+2}} \right\} \quad .$$
$$\tag{21}$$

The simplest way to verify the correctness of this naive supersymmetrization is to write the one-loop two-photon amplitude in the superformalism, and then sewing together the external legs to create an internal photon, using Feynman gauge.

The denominator of eq.( 20) being bosonic, we can again use the proper-time representation eq.( 14) to get it into the exponent, and then absorb this exponent into the worldline superpropagator. The algebra is completely identical to the scalar case, and leads to modified superpropagators $\hat{G}^{(m)}$ which are given by the same formulas as in eqs.( 15) and ( 16), with all the one-loop Green functions appearing on the right-hand sides replaced by the corresponding one-loop superpropagator eq.( 9). The same applies to the determinant factor $(\text{Det}A^{(m)})^{\frac{D}{2}}$, which gives the ratio of the free Gaussian path-integral with $m$ propagator insertions compared to the free one-loop path integral [14].

To summarize, we can obtain a parameter integral representation for the sum of all diagrams with one spinor loop and fixed numbers of photons, N external and m internal, by Wick contracting N vertex operators eq.( 8) with m factors of $\int_0^T d\tau_a d\theta_a \int_0^T d\tau_b d\theta_b DX_a^\mu DX_{b\mu}$, using the modified superpropagator $\hat{G}^{(m)}$.

**5. The 2-Loop QED $\beta$ – function**

As an illustration, we will use this calculus for a rederivation of the two-loop QED $\beta$-function, both for scalar and for spinor electrodynamics. We will work in component formalism for transparency.

As usual, matters much simplify if one is only interested in the $\beta$-function contribution, as opposed to calculation of the whole amplitude. The simplest thing for us to do is to use the effective action formalism with a constant background field $F_{\mu\nu}$, and read off the $\beta$-function from the coefficient of the induced $F_{\mu\nu}F^{\mu\nu}$-term.

As a warm-up, let us first redo the one-loop calculation [5, 7].
For constant $F_{\mu\nu}$, we can choose a gauge such that $A_\mu = \frac{1}{2}x^\rho F_{\rho\mu}$. Using this $A$ – field in the one-loop path-integral eq.( 1) and expanding the interaction exponential to second order, one obtains

$$
\begin{aligned}
\Gamma^{(1)}[F] &= -2 \int_0^\infty \frac{dT}{T} e^{-m^2 T} \int \mathcal{D}x \mathcal{D}\psi \exp\left[-\int_0^T d\tau \left(\frac{1}{4}\dot{x}^2 + \frac{1}{2}\psi\dot{\psi}\right)\right] \\
&\quad \times \left(-\frac{e^2}{2}\right) \int_0^T d\tau_1 \int_0^T d\tau_2 \left[\frac{1}{4}\dot{x}_1^\mu F_{\mu\nu}x_1^\nu \dot{x}_2^\alpha F_{\alpha\beta} x_2^\beta + \psi_1^\mu F_{\mu\nu}\psi_1^\nu \psi_2^\alpha F_{\alpha\beta}\psi_2^\beta\right] \\
&= \frac{e^2}{2} \int_0^\infty \frac{dT}{T}[4\pi T]^{-\frac{D}{2}} e^{-m^2 T} \int_0^T d\tau_1 \int_0^T d\tau_2 \left[\dot{G}_{B12}^2 - G_{F12}^2\right] \int dx_0 F_{\mu\nu}F^{\mu\nu}
\end{aligned}
\tag{22}
$$

(we have abbreviated $\dot{G}_B(\tau_1,\tau_2)$ by $\dot{G}_{B12}$ etc.). The double parameter integral (which is really only a single one, as translation invariance can be used to set $\tau_1 = 0$) gives

$$
\int_0^T d\tau_1 \int_0^T d\tau_2 \left[\dot{G}_B(\tau_1,\tau_2)^2 - G_F(\tau_1,\tau_2)^2\right] = -\frac{2}{3}T^2 \quad . \tag{23}
$$

To extract the divergent part of the remaining global proper-time integral, various regularization methods could be employed. Choosing dimensional regularization, we obtain

$$
\int_0^\infty \frac{dT}{T} e^{-m^2 T} \to \int_0^\infty \frac{dT}{T} e^{-m^2 T} T^{2-\frac{D}{2}} \sim -\frac{2}{\epsilon} \tag{24}
$$

with $\epsilon = D - 4$. Putting things together, the desired one-loop contribution to the effective action becomes

$$
\Gamma^{(1)}[F] \sim \frac{2}{3}(4\pi)^{-2}\frac{1}{\epsilon}e^2 \int dx_0 F_{\mu\nu}F^{\mu\nu} \quad . \tag{25}
$$

From this one obtains the one-loop photon wave-function renormalization factor

$$(Z_3 - 1)^{(1)} = \frac{2}{3\epsilon} \frac{\alpha}{\pi} \quad , \qquad (26)$$

leading to the usual value for the one-loop QED $\beta$ – function,

$$\beta^{(1)}(\alpha) = \frac{2}{3} \frac{\alpha^2}{\pi} \qquad (27)$$

($\alpha = \frac{e^2}{4\pi}$).

Now let us describe the two-loop calculation. In the Feynman diagram calculation (see e.g. [23]), one would have to separately calculate the three diagrams of fig. 1 using some regularization, say dimensional regularization, and then extract their $\frac{1}{\epsilon}$ – poles. Cancellation of the $\frac{1}{\epsilon^2}$ – poles would be found in the sum of the results, indicating a cancellation of subdivergences due to gauge invariance.

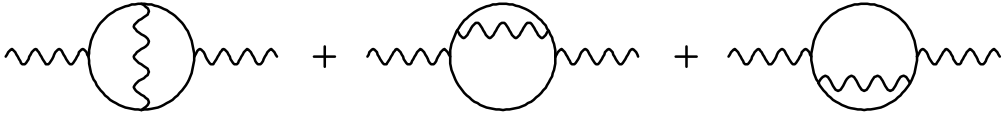

Figure 1: Diagrams contributing to the two-loop vacuum polarization

It will be seen that, in the present formalism, the three diagrams get combined into one calculation.

Let us begin with the purely bosonic contributions, which correspond to the scalar QED calculation. Those are obtained by inserting the worldline current-current interaction term eq.( 19) into the bosonic one-loop path-integral. After exponentiation of the denominator and absorption into the worldline Green function, this results in

$$\begin{aligned}\Gamma^{(2)}_{bos}[F] &= -2\Gamma^{(2)}_{scal}[F] \\ &= -2(4\pi)^{-D} \int_0^\infty \frac{dT}{T} e^{-m^2 T} T^{-\frac{D}{2}} \int_0^\infty d\bar{T} \int_0^T d\tau_a \int_0^T d\tau_b [\bar{T} + G_{Bab}]^{-\frac{D}{2}} \\ &\quad \times \left(\frac{-e^2}{2}\right)^2 \int_0^T d\tau_1 \int_0^T d\tau_2 \int dx_0 \frac{1}{4} \langle \dot{y}_1^\mu F_{\mu\nu} y_1^\nu \dot{y}_2^\alpha F_{\alpha\beta} y_2^\beta \dot{y}_a^\lambda \dot{y}_{b\lambda} \rangle \quad . \quad (28)\end{aligned}$$

Note the appearance of the two-loop determinant factor $[\bar{T} + G_B(\tau_a, \tau_b)]^{-\frac{D}{2}}$. The Wick contraction of

$$\langle \ddot{y}_1^\mu \dot{y}_1^\nu \ddot{y}_2^\alpha \dot{y}_2^\beta \dot{y}_a^\lambda \dot{y}_{b\lambda} \rangle \tag{29}$$

has now to be done, using the two-loop Green function eq.( 15) (care must be taken with Wick contractions involving $\dot{y}_a, \dot{y}_b$, as the derivatives should not act on the $\tau_a, \tau_b$ explicitly appearing in that Green function).

Due to the symmetries of the problem, there are only two nonequivalent contraction possibilities. The result is written out in terms of the bosonic one-loop Green function and its derivatives. As in the one-loop calculation, one next eliminates all factors of $\ddot{G}_B$ appearing by partial integrations with respect to $\tau_1, \tau_2, \tau_a, \tau_b$.

As the next step, all fermionic contributions are included by applying the one-loop substitution rule, replacing, for example,

$$\dot{G}_{B12}\dot{G}_{B21}\dot{G}_{Bab}\dot{G}_{Bba} \to (\dot{G}_{B12}\dot{G}_{B21} - G_{F12}G_{F21})(\dot{G}_{Bab}\dot{G}_{Bba} - G_{Fab}G_{Fba}) \tag{30}$$

etc.

The integrations must then be carried out. At this stage, what we have is the desired contribution to the two-loop effective action in form of a sixfold integral (see fig. 2),

$$\Gamma^{(2)}_{spin}[F] = -2(4\pi)^{-D}\frac{e^4}{16}\int_0^\infty \frac{dT}{T}e^{-m^2 T}T^{-\frac{D}{2}}\int_0^\infty d\bar{T}$$
$$\times \int_0^T d\tau_a d\tau_b d\tau_1 d\tau_2 \, P(T, \bar{T}, \tau_a, \tau_b, \tau_1, \tau_2) F_{\mu\nu} F^{\mu\nu} \quad . \tag{31}$$

The function P (whose bosonic part $P_{bos}$ is given in appendix A) is a polynomial in the various $G_{Bij}, \dot{G}_{Bij}, G_{Fij}$, multiplied by powers of $[\bar{T} + G_B(\tau_a, \tau_b)]^{-1}$. In particular, the integrations over $\tau_1, \tau_2$ are polynomial, and can be performed easily, either by computer, or using a set of relations of the type

$$\int_0^1 du_2 \dot{G}_{12}\dot{G}_{B23} = 2G_{B13} - \frac{1}{3}$$
$$\int_0^1 du_2 G_{B12}G_{B23} = -\frac{1}{6}G_{B13}^2 + \frac{1}{30}$$
$$\vdots \qquad \vdots$$

$$\tag{32}$$

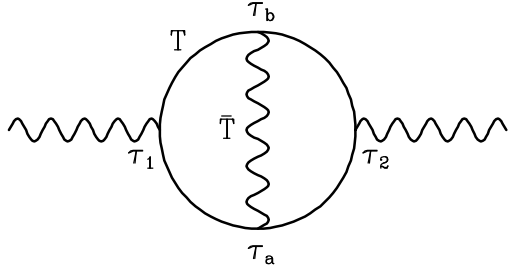

Figure 2: Definition of the six integration parameters

which may be derived from the master identities

$$\int_0^1 du_2 \ldots du_n \dot{G}_{B12} \dot{G}_{B23} \ldots \dot{G}_{Bn(n+1)} = -\frac{2^n}{n!} B_n(|u_1 - u_{n+1}|) \text{sign}^n(u_1 - u_{n+1})$$

$$\int_0^1 du_2 \ldots du_n G_{F12} G_{F23} \ldots G_{Fn(n+1)} = \frac{2^{n-1}}{(n-1)!} E_{n-1}(|u_1 - u_{n+1}|) \text{sign}^n(u_1 - u_{n+1}) \quad .$$

(33)

In writing those identities, we have scaled down to the unit circle again ($\tau_i = T u_i$). $B_n$ denotes the n-th Bernoulli-polynomial, and $E_n$ the n-th Euler-polynomial. Due to the fact that those polynomials can be rewritten as

$$B_n(x) = P_n(x^2 - x) \quad \text{(n even)}$$
$$B_n(x) = P_n(x^2 - x)(x - \frac{1}{2}) \quad \text{(n odd)}$$

(34)

with another set of polynomials $P_n(x)$ (the same property holds true for $E_n(x)$), the right hand sides can always be reexpressed in terms of $G_B, \dot{G}_B$ and $G_F$, so that explicit $u_i$'s will never appear in those relations. Those integrals needed for the present calculation are listed in appendix B (for the spinor-loop case); a proof of the general identities eqs.( 33) is given in appendix C [1].

Next we perform the $\bar{T}$ – integration, which is trivial:

---

[1] For the traces of the left hand sides, recursion relations had already been derived in [6].

$$\int_0^\infty d\bar{T}\,[\bar{T}+G_{Bab}]^{-\frac{D}{2}-k} = \frac{G_{Bab}^{1-\frac{D}{2}-k}}{\frac{D}{2}+k-1} \qquad (k=1,2). \qquad (35)$$

Collecting terms, we get

$$\int_0^\infty d\bar{T}\int_0^T d\tau_1\int_0^T d\tau_2\, P(T,\bar{T},\tau_a,\tau_b,\tau_1,\tau_2) =$$
$$\frac{16}{3D}\Big\{(D-4)(D-1)G_{Bab}^{1-\frac{D}{2}}T + (D-2)(D-7)G_{Bab}^{2-\frac{D}{2}}\Big\} \quad. \qquad (36)$$

Let us also give the corresponding expression for scalar QED, which is obtained by using only the bosonic part $P_{bos}$ of the function $P$:

$$\int_0^\infty d\bar{T}\int_0^T d\tau_1\int_0^T d\tau_2\, P_{bos}(T,\bar{T},\tau_a,\tau_b,\tau_1,\tau_2) = \frac{2}{3}(D-1)G_{Bab}^{-\frac{D}{2}}T^2$$
$$+(D-1)(\frac{32}{3D}-4)G_{Bab}^{1-\frac{D}{2}}T + \frac{16}{3D}(D-2)(D-7)G_{Bab}^{2-\frac{D}{2}} \quad. \qquad (37)$$

Setting $\tau_a = 0$, the integration over $\tau_b$ produces a couple of Euler Beta-functions,

$$\int_0^T d\tau_a\int_0^T d\tau_b\, G_{Bab}^{k-\frac{D}{2}} = B\!\left(k+1-\frac{D}{2},k+1-\frac{D}{2}\right)T^{2+k-\frac{D}{2}} \quad. \qquad (38)$$

As in the one-loop case, the remaining electron proper-time integral just gives a $\Gamma$ – function:

$$\int_0^\infty \frac{dT}{T}e^{-m^2T}T^{4-D} = \Gamma(4-D)m^{2(D-4)} \quad. \qquad (39)$$

Combining terms and performing the $\epsilon$ – expansions for the effective lagrangians, we obtain

$$\mathcal{L}_{scal}^{(2)}[F] \sim \frac{1}{2\epsilon}e^4(4\pi)^{-4}F_{\mu\nu}F^{\mu\nu} + O(\epsilon^0)$$
$$\mathcal{L}_{spin}^{(2)}[F] \sim -\frac{3}{\epsilon}e^4(4\pi)^{-4}F_{\mu\nu}F^{\mu\nu} + O(\epsilon^0) \quad.$$
$$(40)$$

So far this is a calculation of the bare regularized effective action. What about renormalization? The counterdiagrams due to electron wave function and vertex renormalization need not be taken into account, as they cancel by the QED Ward identity ($Z_1 = Z_2$). However, we have used the electron mass as an infrared regulator for the electron proper-time integral eq.( 39); mass renormalization must therefore be dealt with.

Generally, we do not know, at present, how to perform renormalization completely in terms of worldline concepts; we have to refer to standard field theory for this part of the calculation.

Our calculation corresponds to a Feynman calculation in dimensional regularization and Feynman gauge, so we need to know the corresponding one-loop mass renormalization counterterms, both for scalar and spinor QED. This is a simple textbook calculation, of which we give the result only:

$$\begin{aligned} \frac{\delta m_{scal}^2}{m_{scal}^2} &= \frac{6}{\epsilon} e^2 (4\pi)^{-2} \\ \frac{\delta m_{spin}}{m_{spin}} &= \frac{6}{\epsilon} e^2 (4\pi)^{-2} \end{aligned}$$

(41)

Insertions of those counterterms into the one-loop path integral produce the following contributions to the two-loop effective lagrangians,

$$\begin{aligned} \Delta \Gamma_{scal}^{(2)}[F] &= \delta m_{scal}^2 \frac{\partial}{\partial m^2} \Gamma_{scal}^{(1)}[F] \\ &\sim \frac{1}{2\epsilon} e^4 (4\pi)^{-4} \int dx_0 F_{\mu\nu} F^{\mu\nu} + O(\epsilon^0) \\ \Delta \Gamma_{spin}^{(2)}[F] &= \delta m_{spin} \frac{\partial}{\partial m} \Gamma_{spin}^{(1)}[F] \\ &\sim \frac{4}{\epsilon} e^4 (4\pi)^{-4} \int dx_0 F_{\mu\nu} F^{\mu\nu} + O(\epsilon^0) \quad . \end{aligned}$$

(42)

Here $\Gamma_{spin}^{(1)}$ denotes the one-loop path integral eq.( 22), $\Gamma_{scal}^{(1)}$ its scalar QED counterpart.

Extraction of the $\beta$ – function coefficients proceeds in the usual way. From the total effective lagrangians

$$\begin{aligned}\mathcal{L}_{scal}^{(2)}[F] + \Delta\mathcal{L}_{scal}^{(2)}[F] &\sim \frac{1}{\epsilon}e^4(4\pi)^{-4}F_{\mu\nu}F^{\mu\nu} \\ \mathcal{L}_{spin}^{(2)}[F] + \Delta\mathcal{L}_{spin}^{(2)}[F] &\sim \frac{1}{\epsilon}e^4(4\pi)^{-4}F_{\mu\nu}F^{\mu\nu}\end{aligned} \quad (43)$$

one obtains the two-loop photon wave-function renormalization factors, and from those the standard results for the two-loop $\beta$ – function coefficients [24, 25],

$$\beta_{scal}^{(2)}(\alpha) = \beta_{spin}^{(2)}(\alpha) = \frac{\alpha^3}{2\pi^2} \quad . \quad (44)$$

Observe that in the spinor-loop case, the integrand after performance of the first three integrations, eq.( 36), has only one term which would be divergent for $D = 4$ when integrated over $\tau_b$. Moreover, the coefficient of this term vanishes for $D = 4$. This suggests that this calculation can be further simplified by using some four-dimensional regularization scheme. And indeed, if we do the spinor-loop calculation in four dimension, then instead of eq.( 36) we find simply

$$\int_0^\infty d\bar{T} \int_0^T d\tau_1 \int_0^T d\tau_2 \, P(T, \bar{T}, \tau_a, \tau_b, \tau_1, \tau_2) = -8 \quad . \quad (45)$$

This time there is no dependence on $\tau_a, \tau_b$ left, so that one immediately gets

$$\mathcal{L}_{spin}^{\prime(2)}[F] = (4\pi)^{-4}e^4 \int_0^\infty \frac{dT}{T} e^{-m^2 T} F_{\mu\nu}F^{\mu\nu} \quad . \quad (46)$$

It is only the final electron proper-time integral that now needs to be regularized. This can be done by introducing a proper-time cutoff $T_0$ at the lower integration limit, which replaces eq.( 39) by

$$\int_{T_0}^\infty \frac{dT}{T} e^{-m^2 T} \sim -\ln(m^2 T_0) \quad (47)$$

(Pauli-Villars regularization could be used as well, though proper-time regularization appears more natural in the worldline formalism). With this regulator, the two-loop effective lagrangian becomes

$$\mathcal{L}_{spin}^{\prime(2)}[F] \sim -\ln(m^2 T_0)(4\pi)^{-4}e^4 F_{\mu\nu}F^{\mu\nu} + \text{finite} \quad . \quad (48)$$

In spite of the apparent suppression of subdivergences, there is again a contribution from mass renormalization, which can be determined by comparison with the corresponding Feynman calculation [2]. On-shell renormalization of spinor QED using a proper-time cutoff has been studied in refs. [27, 28]. It leads to a one-loop mass renormalization counterterm

$$\frac{\delta m}{m} = 3\ln(m^2 T_0)e^2(4\pi)^{-2} + \text{finite} \quad . \tag{49}$$

Insertion of this counterterm into the one-loop path integral gives

$$\begin{aligned}
\Delta\Gamma'^{(2)}_{spin}[F] &= \delta m \frac{\partial}{\partial m}\Gamma'^{(1)}[F] \\
&\sim 2\ln(m^2 T_0)(4\pi)^{-4}e^4 \int dx_0 F_{\mu\nu}F^{\mu\nu} + \text{finite},
\end{aligned} \tag{50}$$

so that mass renormalization now just amounts to a sign change for the effective lagrangian:

$$\mathcal{L}'^{(2)}_{spin}[F] + \Delta\mathcal{L}'^{(2)}_{spin}[F] \sim \ln(m^2 T_0)(4\pi)^{-4}e^4 F_{\mu\nu}F^{\mu\nu} \quad . \tag{51}$$

The extraction of the (still scheme-independent) $\beta$ – function coefficient $\beta^{(2)}_{spin}(\alpha)$ is again standard [28], and leads back to eq.( 44).

Proper-time regularization could, of course, also be applied to the corresponding scalar QED calculation. However, here one either has to regulate both the electron and the photon proper-time integrals, or to switch to Landau gauge, where the $\tau_b$ – integral again becomes finite in $D = 4$ (as we have verified).

A word of explanation may be in place for our use of the one-loop substitution rule eq.( 30) in the two-loop context. The continued validity of the substitution rule at the multiloop-level is a consequence of the compatibility of the superfield formalism with the mentioned sewing-procedure, as will be explained in more detail elsewhere [32]. As an explicit check, we have performed this calculation in yet another way, namely by writing the one-loop four-photon amplitude in the Bern-Kosower representation, and then sewing together two of the photon legs to create the two-loop vacuum polarization amplitude. Albeit requiring considerably more work, this procedure ultimately yields exactly the same parameter integrals as the use of the two-loop worldline Green function.

Moreover, the calculation in the dimensional scheme has been checked in detail [29] against a Feynman calculation using the second-order Feynman rules of [30].

---

[2]Mass renormalization had been omitted in an earlier, incorrect version of this calculation [26].

Let us summarize the properties of this calculation:

i) Neither momentum integrals nor Dirac traces had to be calculated.

ii) Only simple one-dimensional integrals were encountered.

iii) The three diagrams of fig. 1 were combined into one calculation (in fact, in this formalism it is somewhat *easier* to compute the sum than any single one of them).

iv) In the spinor-loop case, introduction of a regulator could be avoided, except for the final electron proper-time integration.

Property iii) is a general property of the formalism, which should become increasingly important at higher orders. Property iv) may well be accidental to the two-loop case, as far as Feynman gauge is concerned. However, we expect it to hold true in general for the recursively determined gauge where $Z_1 = Z_2 = 1$ [31].

## 6. Conclusions

To conclude, we have shown that the generalization of the Bern-Kosower formalism proposed in [14] holds considerable promise as a tool for multiloop calculations in quantum electrodynamics. It allows to write down compact integral representations combining all Feynman diagrams with one spinor-loop and a fixed number of internal and external photons (the extension to an arbitrary number of spinor-loops is straightforward, as will be shown elsewhere [32]). This kind of sum of diagrams, however, is known to be afflicted with extensive cancellations between diagrams. Those cancellations are clearly related to gauge invariance, and to the fact that the Feynman diagram calculation splits a gauge invariant amplitude into non-gauge invariant pieces. For instance, it is a well-known fact that, at any fixed loop order, the higher-order poles cancel and only the $\frac{1}{\epsilon}$ – pole persists in the sum of all single-spinor loop Feynman diagrams contributing to the vacuum polarization. Moreover, individual Feynman diagrams in this sum contribute transcendental numbers to the $\beta$ – function coefficients, which happen to cancel out for those few coefficients which have been calculated [33, 34]. While the cancellation of higher-order poles is a well-understood consequence of gauge invariance [31], no convincing explanation has been given, so far, for the cancellation of transcendentals [3]. We hope that the formalism developed in this paper will serve to shed new light on this old problem.

*Acknowledgements:*
We would like to thank Z. Bern, D. Broadhurst, M. Jamin, A.L. Kataev, U. Müller, M. Reuter and J.L. Rosner for various discussions and informations. Computer support by P. Haberl is gratefully acknowledged.

---

[3] Very recently, concepts from knot theory have been used to establish a link between both types of cancellations [35].

## Appendix A

We give here the purely bosonic part $P_{bos}$ of the function $P(T,\bar{T},\tau_a,\tau_b,\tau_1,\tau_2)$ appearing in eq. (31). The full $P$ is obtained from this by application of the substitution rule eq.( 11). In writing this polynomial, we have used symmetry with regard to interchange of $\tau_1$ and $\tau_2$ to combine some terms, and omitted some terms which are total derivatives with respect to $\int d\tau_1$ or $\int d\tau_2$ (those terms are easy to identify at an early stage of the calculation).

$$\begin{aligned}
P_{bos} &= \gamma^{\frac{D}{2}}\Big\{12\gamma\dot{G}_{Bab}^2\dot{G}_{B12}^2 + 32\gamma\dot{G}_{Bab}\dot{G}_{B12}\dot{G}_{B1a}\dot{G}_{B2b} \\
&\quad +8\gamma\dot{G}_{B1a}\dot{G}_{Bab}\dot{G}_{B12}[\dot{G}_{B2a}-\dot{G}_{B2b}] \\
&\quad -4\gamma\dot{G}_{B1a}\dot{G}_{B2b}[\dot{G}_{B1a}-\dot{G}_{B1b}][\dot{G}_{B2a}-\dot{G}_{B2b}] \\
&\quad +18\gamma^2\dot{G}_{Bab}^2\dot{G}_{B12}[\dot{G}_{B1a}-\dot{G}_{B1b}][G_{B2a}-G_{B2b}]\Big\} \;.
\end{aligned}$$
(A.1)

We have defined $\gamma = [\bar{T}+G_{Bab}]^{-1}$.

## Appendix B

Integrals occuring in the calculation of the two-loop spinor-QED $\beta$-function:

$$\int_0^1 du_1 \int_0^1 du_2 (\dot{G}_{B12}^2 - G_{F12}^2) = -\frac{2}{3} \quad \text{(B.1)}$$

$$\int_0^1 du_1 [G_{B13}-G_{B14}]^2 = \frac{1}{3}G_{B34}^2 \quad \text{(B.2)}$$

$$\int_0^1 du_1 \int_0^1 du_2 (\dot{G}_{B12}\dot{G}_{B23}\dot{G}_{B34}\dot{G}_{B41}$$
$$-G_{F12}G_{F23}G_{F34}G_{F41}) = -\frac{8}{3}G_{B34}^2 - \frac{4}{3}G_{B34} \quad \text{(B.3)}$$

$$\int_0^1 du_1 \int_0^1 du_2 (\dot{G}_{B13}\dot{G}_{B32}\dot{G}_{B24}\dot{G}_{B41}$$
$$-G_{F13}G_{F32}G_{F24}G_{F41}) = 4G_{B34}^2 + \frac{8}{3}G_{B34} - \frac{8}{9} \quad \text{(B.4)}$$

# Appendix C

To prove the first of the identities in eq. (33), observe that, by construction, $\frac{1}{2}\dot{G}_B$ is the integral kernel inverting the first derivative $\partial_B$ acting on periodic functions. We may therefore write

$$\begin{aligned} K_n(u_1 - u_{n+1}) &:= \int_0^1 du_2 \ldots du_n \dot{G}_{B12}\dot{G}_{B23}\ldots \dot{G}_{Bn(n+1)} \\ &= 2^n < u_1 \mid \partial_B^{-n} \mid u_{n+1} > . \end{aligned} \tag{C.1}$$

This leads to the recursion relation

$$\frac{\partial}{\partial u} K_n(u - u') = 2^n < u \mid \partial_B^{-(n-1)} \mid u' > = 2 K_{n-1}(u - u'). \tag{C.2}$$

We want to show that the same recursion relation is fulfilled by the polynomial $\tilde{K}_n$,

$$\tilde{K}_n(u - u') := -\frac{2^n}{n!} B_n(|u - u'|)\operatorname{sign}^n(u - u'). \tag{C.3}$$

Explicit differentiation yields

$$\begin{aligned} \frac{\partial}{\partial u} \tilde{K}_n(u - u') &= -\frac{2^n}{n!} B'_n(|u - u'|)\operatorname{sign}(u - u')\operatorname{sign}^n(u - u') \\ &= -\frac{2^n}{(n-1)!} B_{n-1}(|u - u'|)\operatorname{sign}^{n+1}(u - u') \\ &= 2\tilde{K}_{n-1}(u - u'). \end{aligned} \tag{C.4}$$

Here the recursion relation for the Bernoulli polynomials was used, $B'_n(x) = nB_{n-1}(x)$. An additional term arising by differentiation of the signum function for $n$ odd can be deleted due to the fact that

$$\delta(x)B_n(\mid x \mid) = \delta(x)B_n(0) = 0 \tag{C.5}$$

for $n$ odd, $n > 1$. The proof is completed by checking that the master identity works for $n = 1$ ($B_1(x) = x - \frac{1}{2}$), and on the diagonal $u_1 = u_{n+1}$ for any $n$ (this special case has already been proven in [7]).
The proof of the second master identity proceeds in a completely analogous way.